\documentclass[11pt]{article}
\usepackage{vietnam}

\usepackage{color}
\usepackage{xcolor}
\usepackage{subfig}
\usepackage{caption}
\usepackage{graphicx}
\usepackage{float}
\usepackage{pgfplots}
\usepackage{tikz}
\usetikzlibrary{arrows,shapes,positioning}
\usetikzlibrary{decorations.markings,decorations.pathmorphing,decorations.pathreplacing}
\usetikzlibrary{calc,patterns,shapes.geometric}
\usepgfplotslibrary{groupplots}
\usepackage{mdframed}
\usetikzlibrary{patterns}
\pgfplotsset{compat=1.5}
\usepackage{wrapfig}
\usepackage{rotating}
\usepackage{lscape}
\usepackage{multirow}
\usepackage{threeparttable}
\usepackage{fancybox}
\usepackage{blindtext}
\usepackage{verbatim}
\usepackage{algpseudocode}
\algtext*{EndWhile}% Remove "end while" text
\algtext*{EndIf}% Remove "end if" text
\usepackage{listings}	
\usepackage{amsmath,amssymb,amsfonts,amsthm}
\usepackage{bm}% bold math
\usepackage{mathrsfs}
\usepackage{tensor}
\graphicspath{{./images/}}
\usetikzlibrary{spy}
\usepackage{bm}

\def\qu{\mathrm{qu}}
\def\cl{\mathrm{cl}}

\def\T{\mathcal{T}}

\def\age{\text{age}}
\newcommand\subrel[2]{\mathrel{\mathop{#2}\limits_{#1}}}

\newcommand{\Photo}{}

\begin{document}
\vspace*{4cm}
\title{QUANTUM SPINDOWN OF HIGHLY MAGNETIZED NEUTRON STARS}

\author{ B. LAMINE, C. BERTHIERE, A. DUPAYS }

\address{IRAP, UPS, 14 avenue Edouard Belin,\\
31450 Toulouse, France}

\maketitle\abstracts{Pulsars are highly magnetized and rapidly
  rotating neutron stars. The magnetic field can reach the critical
  magnetic field from which quantum effects of the vacuum becomes
  relevant, giving rise to magnetooptic properties of vacuum
  characterized as an effective non linear medium. One spectacular
  consequence of this prediction is a macroscopic friction that leads
  to an additionnal contribution in the spindown of pulsars. In this
  paper, we highlight some observational consequences and in
  particular derive new constraints on the parameters of the Crab
  pulsar and J0540-6919.}

\section{Introduction}

It is known from long time that quantum effects give to vacuum its own
electromagnetic properties, in analogy with a usual ponderable media
(for a recent review, see for
example~\cite{2013RPPh...76a6401B}). Among experimentally accessible
consequences are photon/photon scattering~\cite{2013arXiv1305.7142D}
or vacuum birefringence~\cite{2012PhRvA..85a3837B}, both being
targeted by laboratory experiments. Astrophyscial objects such as
Neutron Stars (NS) are also a laboratory to test those quantum
corrections, simply because they sustain a very high magnetic field
that enhances the quantum features of vacuum. In particular, when the
magnetic energy around a NS is comparable to the electron rest energy,
$B\sim B_c=\frac{m_e^2c^2}{e\hbar}$, a significant magnetization arise
in the vacuum that give rise to spectacular macroscopic consequences
on the spindown of this
NS~\cite{2008EL.....8269002D,2012EL.....9849001D}.

The physical principle of the quantum contribution to the spindown is
rather simple. The extremely high magnetic field, generated by the
rotating dipole $\bm{m}$, creates a time-dependent magnetization in
the vacuum around the NS. In return, this magnetization creates a
magnetic field $\bm{B}_\qu$ which feeds back on the NS magnetic
dipole. Due to retardation effect (finite speed of light), this
back-action leads to a torque $\bm{m}\times\bm{B}_\qu$ that slows down
the spinning of the NS. It has been shown that the energy loss rate of
an isolated pulsar via the previous quantum channel, in the limit
$B_0\ll B_c$, is given
by~\cite{2008EL.....8269002D,2012EL.....9849001D}
\begin{equation}
  \label{eq:3}
  \dot{E}_\qu=-\alpha\left(\frac{3\pi^2}{4}\right)\frac{\sin^2{\theta}}{B_c^2\mu_0c}
\frac{B_0^4R^4}{P^2}\;,
\end{equation}
where $\alpha$ is the fine structure constant, $\theta$ the
inclination angle, $B_0$ the surface magnetic field, $c$ the speed of
light, $R$ the radius of the NS and $P=2\pi/\Omega$ the rotation
period. This contribution is an additional one compared to the
classical spindown, which scales differently with respect to the
physical parameters of the NS. Within the vacuum model (oblique
rotator in vacuum), the energy loss rate is given by the usual dipole
formula
\begin{equation}
  \label{eq:4}
  \dot{E}_{\cl}=-\frac{128\pi^5}{3}\frac{B_0^2 \sin^2\theta R^6}{P^4\mu_0c^3}\;,
\end{equation}
where we can see that the classical contribution scales as $P^{-4}$
instead of the $P^{-2}$ for the quantum one. Both contributions are
roughly of the same order when
$\frac{B_0}{B_c}\sim\frac{R\Omega}{c}\frac{1}{\sqrt{\alpha}}
\frac{1}{\sin\theta}$. Therefore, even if the magnetic field is much
smaller than the critical magnetic field, the quantum contribution
will dominate the classical one for large period P (or low
$\Omega$). Hence, the late-time evolution of the spindown of a pulsar
should generically be quantum-dominated, because the period gradually
increases as the NS loses energy. Of course, this conclusion holds
only in the simple model considered here, which is certainly
incomplete. Among the physical phenomena that could change the
previous statement are for instance the inclusion of a real
magnetosphere, or taking into account an alignment of the NS (ie
$\sin\theta\rightarrow0$ as times passes~\cite{2010MNRAS.402.1317Y}).

\section{Quantum spindown}

Using equations (\ref{eq:3})-(\ref{eq:4}) and assuming the total
energy is only rotational kinetic energy $E=\frac{1}{2}J\Omega^2$ ($J$
being the inertia moment), one gets a new evolution equation for the
rotation period,
\begin{equation}
  \label{eq:5}
\dot{E}=\dot{E}_\cl+\dot{E}_\qu\quad\Rightarrow\quad  
\dot{P}=\frac{\T_\cl}{P}+\frac{P}{\T_\qu}\:,
\end{equation}
where the constants $\T_\cl$ and $\T_\qu$ are easily obtained, 
\begin{eqnarray}
  \label{eq:6}
    \T_\cl&\simeq&8.8\times10^{-16}\,\text{s}\left(\frac{B_0}{10^{8}
\,\text{T}}\right)^2 
 \sin^2\theta\left(\frac{R}{10\,\text{km}}\right)^4\left(\frac{1.4\,
\text{M}_\odot}{M}\right)\;;\\
 \T_\qu&\simeq&2.1\times10^{13}\,\text{s}\,\left(\frac{10^{8}
\,\text{T}}{B_0}\right)^4 
\frac{1}{\sin^2\theta}\left(\frac{10\,\text{km}}{R}\right)^2
\left(\frac{M}{1.4\,\text{M}_\odot}\right)\;.
\end{eqnarray}

It is clear that the quantum contribution will be dominant once
$P>\T\equiv\sqrt{\T_\cl\T_\qu}$, with $\T
\simeq140\,\text{ms}\left(\frac{10^{8} \,\text{T}}{B_0}\right)
\left(\frac{R}{10\,\text{km}}\right)$ a characteristic time. From the
measurement of $P_0$, $P_1$ and $n_0$, respectively the present
period, its first derivative and the braking index
($n\equiv2-P\ddot{P}/\dot{P}^2$), it is possible to solve equation
(\ref{eq:5}) and determine $\T_\cl$ and $\T_\qu$,
\begin{equation}
\label{eq:2}
P(t)=\T\left[\left(1+\frac{P_0^2}{\T^2}\right)e^{\frac{t}{\T_\qu}}- 1\right]^{1/2}\quad,\quad
\T_\cl=P_0P_1\frac{n_0-1}{2}\quad,\quad
\T_\qu=\frac{P_0}{P_1}\frac{2}{3-n_0}\;.
\end{equation}

\begin{figure}
  \centering
\includegraphics[width=.5\textwidth]{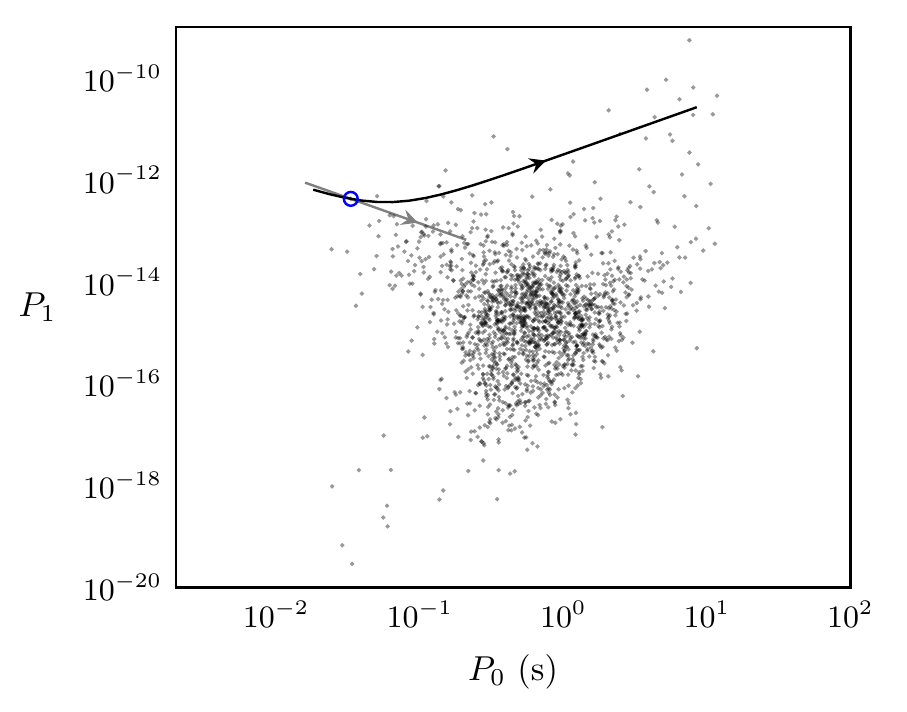}  
\caption{Evolution of the Crab pulsar. The grey line is the classical
  evolution while the black line is the evolution taking into account
  the quantum correction.}
\label{fig:PPdot}
\end{figure}

As an explicit example, the evolution of the Crab pulsar from the
previous expressions is represented in the $(P,\dot{P})$ diagramm of
figure~\ref{fig:PPdot}. The evolution starts from the birth of the
Crab ($~955$ years ago) and last $50\,$kyr. Each dot in this plot is a
pulsar from the ATNF catalog~\cite{2005AJ....129.1993M}. The
interesting feature is that the evolution naturally brings the Crab
towards the so-called magnetar region in the upper right corner, even
if the magnetic field is smaller than the critical field (see next
section for an estimation of the Crab magnetic field, being a few
$10^8\,$T).  This could support the idea already proposed
in~\cite{2012EL.....9849001D} that some of the so-called magnetars are
in fact normal evolved pulsars. A deeper analysis of this hypothesis
is underway.  

The quantum evolution given by (\ref{eq:2}) also has a consequence on
the age of the pulsar, obtained as
\begin{equation}
  \label{eq:8}
  t_\age=\frac{P_0}{P_1}\frac{1}{n_0-3}\,\ln\frac{n_0-1+(3-n_0) 
\frac{P_i^2}{P_0^2}}{2}\subrel{P_i\ll
    P_0}{\sim}\frac{P_0}{(3-n_0)P_1}
\ln\frac{n_0-1}{2}
\end{equation}

The last equality assumes that the present period $P_0$ is much
greater than the initial period $P_i$. This age is always greater than
the characteristic age $\frac{P_0}{2P_1}$ obtained classically with
the same approximation $P_i\ll P_0$. This consequence could be an
observational signature of the quantum evolution since it would show
up as a systematic bias between the kinematical age (or SNR age) and
the characteristic age. Such discrepancies are for example reported
in~\cite{McLaughlin:2002vh}.

% Note also that instead
% of considering the characteristic age, one could also estimate
% classically the age through a constant breaking index spindown
% model. It would then give $\frac{P_0}{P_1(n_0-1)}$, which this time
% would be always greater than the quantum spindown age. Further
% investigations along this direction is ongoing.

\section{Constraints on the mass and radius}

From $\T_\cl$ and $\T_\qu$, it is straightforward to determine $B_0$
and $\sin\theta$ as a function of the mass $M$ and the radius $R$ of
the NS, giving
\begin{eqnarray}
B_0 &=& \frac{5\pi B_cR}{2c\sqrt{\alpha}}\frac{1}{P_0}
\sqrt{\frac{3-n_0}{n_0-1}} \;;\\
  \sin^2\theta&=&\frac{3\alpha}{400\pi^5}\frac{\mu_0
Jc^5}{R^8B_c^2}P_0^3P_1\frac{(n_0-1)^2}{3-n_0}\;.
\end{eqnarray}

\begin{figure}
  \centering
\includegraphics[width=.95\textwidth]{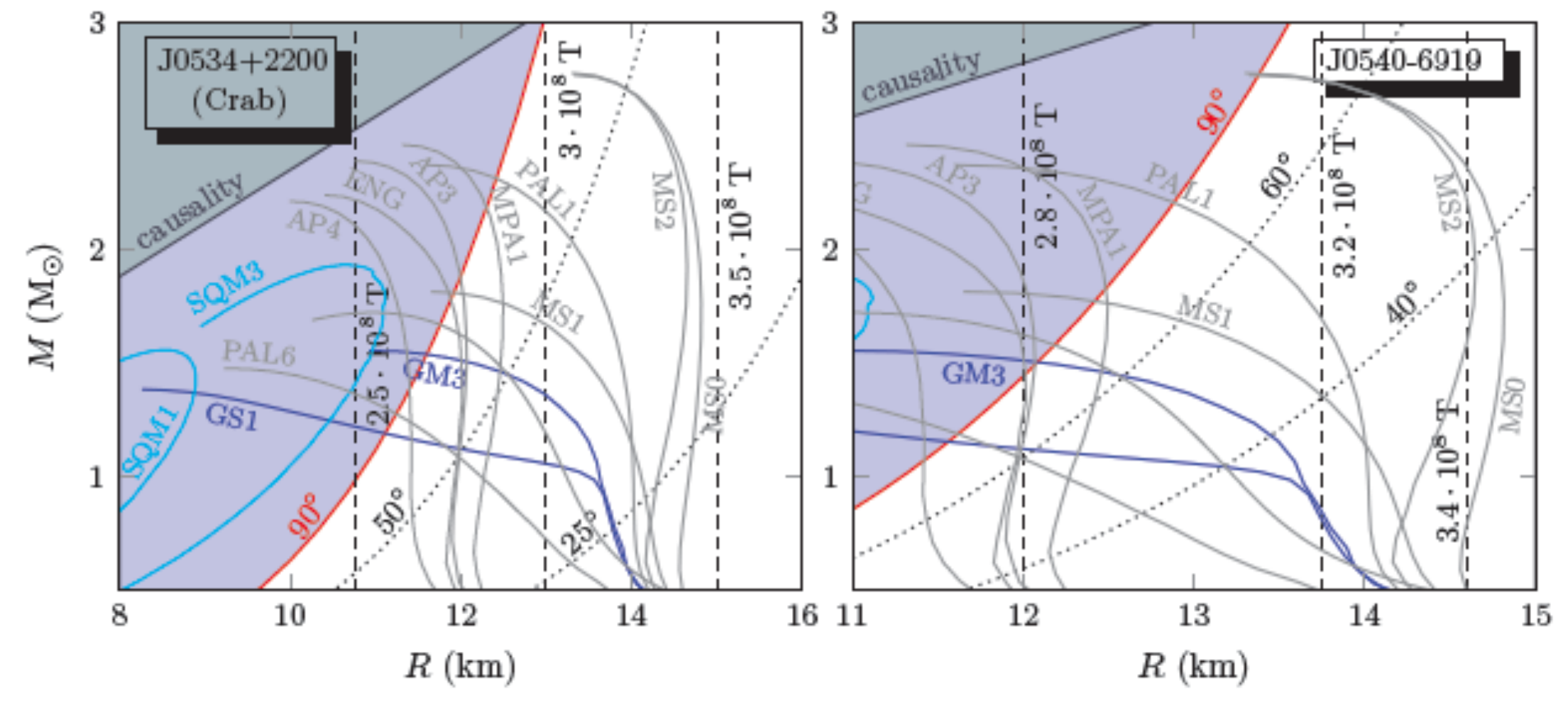}  
\caption{Constraints in the $(M,R)$ diagramm. The colored regions are
  excluded, either by causality or by the condition
  $\sin\theta<1$. The solid lines corresponds to different models of NS
  equation of state. The dotted lines correspond to constant value of the
  inclination angle while vertical dashed lines correspond to constant
  magnetic field $B_0$.}
\label{fig:MR}
\end{figure}

The condition $\sin^2\theta<1$ then provides constraints on the mass
$M$ and the radius $R$ of the pulsar. Such constraints are represented
in the figure~\ref{fig:MR} as exclusion regions, for two similar
pulsars, namely $J0534+2200$ (the Crab) and $J0540-6919$; for those
pulsars the braking index $n_0$ is confidently measured and is given
in table \ref{table:n}. For sake of
simplicity we assumed $J=\frac{2}{5}MR^2$. In the same figure are
represented some families of Equation Of State for the NS (extracted
from~\cite{2007Ap&SS.308..371L}). It is quite remarkable that taking
into account the quantum effect sets some constraints on such EOS. In
particular, the strange quark models (SQM) seem excluded. Of course,
it is not possible to draw any definite conclusion unless a more
realistic model is studied. For example, it is expected that the
magnetosphere could significantly change the previous conclusions.

\begin{table}[h!]
\caption{Spin and breaking index.}
\label{table:n}
\vspace{0.4cm}
  \centering
\begin{tabular}{lccc}
 \hline 
 Name&$n_0$&$P_0$&$P_1$\\
J2000&&(ms)&($10^{-13}$)\\
\hline\hline 
J0534+2200 (Crab) & 2.51 & 33.1 & 4.23 \\
J0540-6919 & 2.14 & 50.5 & 4.79 \\
\hline 
\end{tabular}
\end{table}

\section{Conclusion}

We showed that the predicted quantum-induced spindown in NS leads to
observational consequences that should be looked for carefully. In
particular, the evolution of a pulsar in the $(P,\dot{P})$ diagramm is
qualitatively changed for highly manetized pulsars, the true age of a
pulsar significantly differs from the characteristic age, and some
constraints on the equation of state can be obtained, through new
relationships between the mass, the radius, the inclination angle and
the magnetic field of the NS.

\section*{References}

\bibliography{biblio_pulsars.bib}

% \begin{thebibliography}{99}
% \bibitem{ja}C Jarlskog in {\em CP Violation}, ed. C Jarlskog
% (World Scientific, Singapore, 1988).

% \bibitem{ma}L. Maiani, \Journal{\PLB}{62}{183}{1976}.

% \bibitem{bu}J.D. Bjorken and I. Dunietz, \Journal{\PRD}{36}{2109}{1987}.

% \bibitem{bd}C.D. Buchanan {\it et al}, \Journal{\PRD}{45}{4088}{1992}.

% \end{thebibliography}

\end{document}